\documentclass{article}
\usepackage{frascatiphys,here,graphicx,subfigure,citesort}

\begin{document}
\title{ THE EFFECT OF ISOSPIN VIOLATION ON SCALAR MESON PRODUCTION
}
\author{
 C. Hanhart$^1$, B. Kubis$^2$, and J. R. Pel\'aez$^3$ \\
{$^1$\em Instit\"ut f\"ur Kernphysik (Theorie), Forschungzentrum J\"ulich,} \\{\em D-52425 J\"ulich, Germany} \\
{$^2$\em HISKP (Theorie),
             Universit\"at Bonn, Nussallee 14-16,}\\
{\em  D-53115 Bonn, Germany.} \\
{$^3$\em Departamento de F\'{\i}sica Te\'orica II, Universidad Complutense,}\\{\em E-28040 Madrid. Spain} 
}
\maketitle
\baselineskip=11.6pt
\begin{abstract}
We investigate the isospin-violating
mixing of the light scalar mesons $a_0(980)$ and $f_0(980)$ 
within the unitarized chiral approach. Isospin-violating
effects are considered to leading order in the quark mass difference and
electromagnetism.  In this approach both resonances are generated through
meson-meson dynamics.  
Our results provide a description of the mixing phenomenon 
within a framework consistent with chiral symmetry
and unitarity, where these resonances are not predominantly $q\bar{q}$ states.
We discuss in detail the reactions $J/\Psi\to \phi S$, where $S$ denotes 
a suitable pair of pseudo--scalar mesons in the scalar channel, namely
$\pi^0\eta$, $K^+K^-$, and $K^0 \bar K^0$. In this work predictions
for the cross section in the kaon channels are given for the first time
with isospin violating effects included. 
\end{abstract}
\baselineskip=14pt
\section{Introduction}

Although the light scalar mesons $a_0(980)$ and $f_0(980)$ have been
established as resonances long ago, there is still a heated debate
going on in the literature regarding the very nature of these states.
Naively one might assign them a conventional $q\bar q$ structure,
however, at present no quark model is capable of describing both
states simultaneously as $q\bar q$ states --- see, e.g.,
Ref.\cite{bonn}.  On the other hand, as early as 1977 it was stressed
that especially in the scalar channel the interaction of four-quark
systems (two quarks, two antiquarks) is attractive\cite{jaffe}. Some
authors have found indications for the existence of compact four-quark
states\cite{achasov1,giacosa,vijande}. However, the same short-ranged
interaction can also be the kernel to the scattering of pseudoscalars,
giving rise to extended four-quark states that one might call hadronic
molecules or extraordinary
hadrons\cite{jaffenew1,largeNc1,largeNc2,largeNc3}. Independently, a
similar conclusion was found in different
approaches\cite{isgur,janssen,eef,Oller:1997ti}.

In Refs.\cite{wein,evidence} it is stressed
that the effective coupling constants of the scalar mesons
to the $K\bar K$ channel contain the essential structure
information. Especially, the larger the molecular component, 
the larger the residue at the resonance pole, which acquires
its maximum value in case of a pure molecule.
It should be stressed, however, that this connection can be
made rigorous only for stable bound states and if the bound state pole is on the 
first sheet very close to the elastic threshold\cite{wein}. However,
if the state of interest is narrow and the inelastic threshold
is sufficiently far away, the argument should still hold\cite{evidence}.
Note that both conditions apply for $a_0(980)$ and $f_0(980)$.
Therefore one should aim at observables that are very sensitive
to the effective coupling constants. The resonance signal, as
seen in production experiments for both states, is not 
very useful to determine those couplings, for it turned out
to be mainly sensitive to ratios of couplings\cite{flatte}.
It is therefore important to investigate other observables.

When this formalism was applied to the scalar mesons 
$a_0(980)$ and $f_0(980)$ it was found from an
analysis of a series of reactions\cite{ourscalars} that the latter
is indeed predominantly a $K\bar K$ molecule, in line
with the results of Refs.\cite{isgur,janssen,Oller:1997ti,largeNc1}, while the
results for the former did not lead to  an unambiguous
interpretation. This might either point at a prominent
non--molecular contribution to the $a_0$ structure, or
the $a_0$ is not a bound state, but a virtual state.
To decide on this issue it is important to collect more
information especially on the $a_0$. In Ref.\cite{achasovmix}
it was stressed that the amplitude for the isospin violating
$a_0-f_0$ transition should be very sensitive to the
product of the effective coupling to the two--kaon channels
 of the $f_0$ and the $a_0$. If we assume that the $f_0$ is
a molecule, its coupling to the $K\bar K$ channel is fixed, as
discussed above. Then the value for the $a_0-f_0$ mixing amplitude
should be very sensitive to the structure of the $a_0$.

\begin{figure}
\begin{center}
\includegraphics[width=0.6\linewidth]{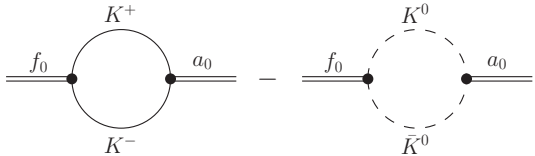}
\end{center}
\caption{\label{lcsb} 
Graphical illustration of the leading contribution to the $a_0-f_0$ mixing
matrix element, driven by the kaon mass differences.
}\end{figure}

\begin{figure}
\begin{center}
\includegraphics[width=0.6\linewidth]{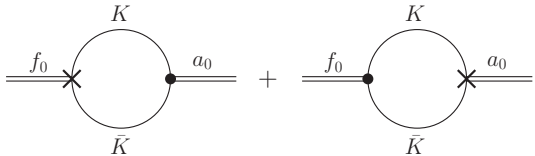}
\end{center}
\caption{\label{nlcsb} 
Graphical illustration of the subleading contribution to the $a_0-f_0$ mixing
matrix element, driven by isospin-violating vertices,
denoted by the crosses.
}\end{figure}

The reason why the mixing amplitude of $a_0$ and $f_0$ is sensitive to
the effective couplings is that it gets a prominent contribution from
the kaon--loops (see Fig.~\ref{lcsb}). Their isospin violating part is
driven by the kaon mass differences giving rise to an effect that can
be shown to scale as $\sqrt{(m_d-m_u)/m_s}$. In contrast to this,
normal isospin violating effects should be of order $(m_d-m_u)/m_s$
--- those could be parameterized, e.g., by isospin--violating coupling
constants (see Fig.~\ref{nlcsb}).  In Ref.\cite{unsers} the latter
effects were calculated for the first time as well to provide a better
quantitative estimate of the mixing amplitude. For this a particular
model needs to be used. We chose the chiral unitary approach which is
a special unitarization procedure for amplitudes calculated using
chiral perturbation theory. The latter provides a systematic inclusion
of quark mass effects while the former allows for an extension of the
formalism into the resonance region --- see Ref.\cite{joseref} for a
recent review.  In addition we also include the isospin breaking
effect of the soft--photon exchange.

In this work we investigate different final states for the
reactions $J/\Psi\to \phi S$~\footnote{This decay was
identified as very useful to study isospin violation for scalar
mesons in Refs.\cite{close,zou}.}, where $S$ denotes 
a suitable pair of pseudo--scalar mesons in the scalar channel, namely
$\pi^0\eta$, $K^+K^-$, and $K^0 \bar K^0$ --- predictions
for the cross section in the kaon channels are given for the first time
with isospin violation included. 
The latter two channels are expected to constrain the charge dependence
of the coupling of the $f_0$ to kaons and should therefore provide
an independent cross check for the size of corrections of order $(m_d-m_u)/m_s$.
In the next section the formalism is briefly reviewed and
the results are given. We close with a short summary.

\section{Formalism and Results}

Details on the formalism are given in Ref.\cite{unsers}. Thus, here we
will only repeat the essential physics that went into the
calculations.  In the Standard Model there are two sources of isospin
violation present, namely the up--down quark mass difference as well
as electromagnetism. Both appear in a well defined form in the
corresponding Lagrangian density formulated in terms of the fundamental
degrees of freedom, here photons, gluons, and quarks. Chiral
perturbation theory\cite{chpt,urech} allows for a consistent
representation of those terms for a theory describing the interaction
of the (pseudo) Goldstone bosons with each other. At leading order in
the chiral Lagrangian the only parameter that appears, in addition to
the various particle masses, is the pion decay constant.  Thus, also
the leading effect of isospin violation is predicted.  It should be
stressed that already at leading order both isospin violating mass
differences as well as isospin violating interactions are present.

\begin{figure}
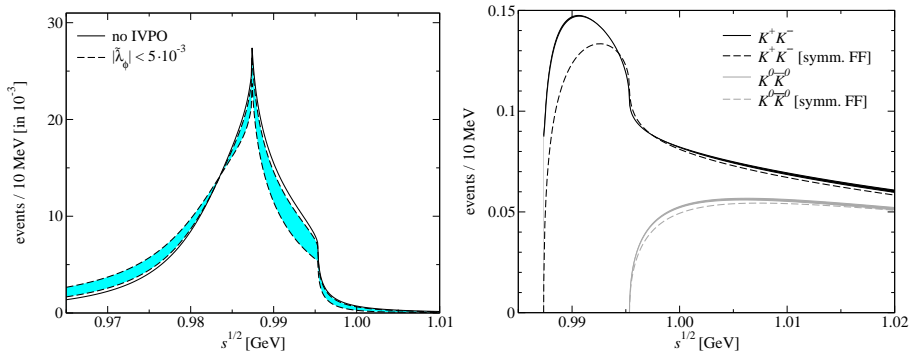

         {\includegraphics[width=0.495\linewidth]{CountRatepietaB_bw.eps} \hfill
          \includegraphics[width=0.495\linewidth]{KKbarSignal_bw.eps}}
\caption{Left: predicted signal for $J/\Psi \to \phi \pi^0\eta$.
The solid line shows the result without isospin violation in the production
operator (IVPO), while the band shows the effect of its inclusion.
Right: predicted signal for $J/\Psi \to \phi K\bar K$. For both kaon
channels the solid band, which includes the uncertainties,
 is the full result, while for the dashed line the same decay amplitude was used
in both channels (see text). }
 \label{Fig:1}
 \end{figure}

Already in isospin symmetric calculations of meson--meson scattering amplitudes, when
unitarizing the leading chiral Lagrangian, an a priori unknown constant
needs to be adjusted to the data.  In addition, a few more
parameters need to be fitted to the production data --- in our case
they were fixed from a fit to the data on $J/\Psi \to \phi \pi^+\pi^-,
K^+K^-$ as well as $J/\Psi \to \omega \pi^+\pi^-$\cite{timo}.  Once
this is done, the isospin violating signals emerge as predictions.
Our results for the various channels are shown in Fig. \ref{Fig:1}.
The predicted signal in the $\pi^0\eta$ channel was already published
in Ref.\cite{unsers}. Note that in an isospin symmetric world this
reaction would not be allowed, thus any signal is directly
proportional to the square of an isospin violating amplitude. Based on
very general scale arguments one can show that in addition this
amplitude is dominated by the isospin violation that emerges from
$a_0-f_0$ mixing\cite{report}.  This was confirmed on the grounds of
the current model: the filled band in the left panel of Fig.
\ref{Fig:1} shows the effect of possible isospin violation in the
production operator.  The right panel of the same figure shows the
$J/\Psi$ decay with a pair of kaons in the final state. Please note
that for these two channels there would be a decay rate also in an
isospin symmetric world, however, then both signals would agree.  The
solid lines show our results for the two channels including the
effects of isospin violation as described in Ref.\cite{unsers}. In
addition, the presence of charged particles in the final state called
for an additional treatment of soft photons. Here we followed
Ref.\cite{gino}.  The decay spectrum for $J/\Psi \to \phi K^+K^-$,
based on an isospin symmetric calculation, is shown in
Ref.\cite{timo}.

 To see how much of the difference in the two kaon channels originates
 from the different phase space thresholds alone ($2m_{K^+}-2m_{K^0}=8$~MeV), in the
 figure we also show the signals that emerge
 when the same decay amplitude is used for both channels (dashed lines). For this
 calculation we took the average of the charged and the neutral kaon
 amplitudes, corresponding to a formal isospin 0 combination.
Obviously the by far most important difference between
the channels is driven by the displacement of the thresholds,
the effect of which is further enhanced by the strongly varying
amplitudes precisely in this threshold region.
The original idea was to extract information on the charge
dependence of the couplings of the $f_0$ to kaons from a comparison
of $J/\Psi \to \phi K^+K^-$ and  $J/\Psi \to \phi K^0 \bar K^0$.
However, as can be seen from the figure, 
the differences between the solid and the dashed lines are too small
to be accessible experimentally.  Therefore we do not expect data for
the kaon channels to be sufficiently sensitive to extract isospin
violating effects in the decay amplitudes.

\section{Summary}

In this work we calculated the reactions  $J/\Psi\to \phi S$, where $S$ denotes 
a suitable pair of pseudo--scalar mesons in the scalar channel, namely
$\pi^0\eta$, $K^+K^-$, and $K^0 \bar K^0$. The goal of this study was
to gain a better quantitative understanding of the phenomenon of $a_0-f_0$ mixing,
which should eventually reveal important information on the structure
of the $a_0(980)$. In addition to the $\pi^0\eta$ channel, the kaon channels
including isospin violation were discussed here for the first time. We
found that, at least within the current model, the impact of the charge
dependence of the coupling of the $f_0$ to kaons is too small to be deduced
from a comparison of the rates for $K^+K^-$, and $K^0 \bar K^0$. 
On the other hand the $\pi^0\eta$ decay channel appears to be not only very sensitive
to the effective coupling constants that encode the structure information but
also theoretically under control\cite{unsers}. For a discussion of additional
background effects see Ref.\cite{zou}.
The corresponding measurements can be performed, once BES III is in operation.

\section{Acknowledgments}

The authors acknowledge partial financial support by the EU integrated
infrastructure initiative HADRONPHYSICS PROJECT,
under contract RII3-CT-2004-506078. 
B.K.\ is supported by the DFG (SFB/TR 16).
The research of J.R.P.\ is partially funded by Banco Santander/Complutense
contract PR27/05-13955-BSCH and Spanish CICYT contracts
FPA2007-29115-E, FPA2005-02327, BFM2003-00856.

\end{document}